\documentstyle[prl,aps,twocolumn]{revtex}
\begin{document}
\draft
\flushbottom
\title{Excitation spectrum of the two-dimensional attractive Hubbard model}
\author{Massimiliano Capezzali and Hans Beck}
\address{Physics Department,
University of Neuch\^{a}tel, \\
Rue A.L. Breguet 1, 
2000 \underline{Neuch\^{a}tel},
CH-Switzerland}
\maketitle
\begin{abstract}
\begin{center}
\parbox
{8cm}
{
We calculate the one-particle spectral functions above the superconducting
transition temperature $T_{C}$, in the framework of a functional
integral approach. The coupling of the electronic self-energy to pair
fluctuations, which are treated by means of a time-dependent
Ginzburg-Landau equation, yields a double-peak structure, around the
Fermi wavenumber. The peak separation is essentially
temperature-independent, but the structure sharpens when $T_{C}$
is approached.
}
\end{center}
\end{abstract}
\pacs{}
\vspace*{-1.0truecm}
\noindent
Although the two-dimensional attractive Hubbard model does not represent
a suitable model to describe the complex physical behavior of the
high-temperature superconducting compounds, it has recently attracted renewed
interest \cite{New1,New2,New3},
on one hand because of its simplicity and, on the other hand,
because it incorporates two fundamental properties of the cuprate
superconductors : working in two dimensions permits to take into
account the strong anisotropy of the latter materials
(at least in the underdoped regime), while
the on-site attraction can account for the characteristic
short in-plane coherence
length, which has been experimentally determined.\\
In a previous paper \cite{Max1}, we calculated the one-electron Green's
function of the two-dimensional attractive Hubbard model, by coupling
the charge carriers to pair fluctuations, which were subsequently
approximated by a homogeneous amplitude and by phase fluctuations,
corresponding to the 2D-XY model; we found that the electronic density
of states shows a pseudogap at temperatures well above the superconducting
transition temperature $T_{C}$. \\
In the present communication, we shall treat
pair fluctuations above $T_{C}$ by means of a time-dependent Ginzburg-
Landau equation. In order to derive the latter, the interaction
term appearing in the Hamiltonian of
the attractive Hubbard model is decoupled by a standard Stratonovich-Hubbard
transformation, through which a complex, fluctuating pairing field
$\Delta$ appears.
The partition function is thereafter written as the functional integral,
over the latter field, of an effective action which, upon exploiting
the usual cumulant expansion for expectation values up to fourth order
in $\Delta$,
becomes a dynamic Ginzburg-Landau-like functional, $S(\Delta)$.
The expectation value
of the squared modulus of the pairing field $\Delta$ has the form :
\begin{eqnarray}\label{e1}
\Xi(\vec{q},z_{\alpha})\equiv
\left<\mid\Delta(\vec{q},z_{\alpha})\mid^{2}\right>={1\over a
+cq^{2}+(d_{1}+id_{2})z_{\alpha}},
\end{eqnarray}
where $a$, $c$, $d_{1}$ and $d_{2}$ are real coefficients
that emerge from expanding the inverse of the
two-particle propagator, for
small wavenumbers $\vec{q}$ and (bosonic Matsubara) frequencies $z_{\alpha}$.
To lowest order, the self-energy is given by \cite{Max1} :
\begin{eqnarray}\label{e2}
\sigma(\vec{k},z_{\nu})=-\sum_{\vec{q},z_{\alpha}}^{}{
\left<\mid\Delta(\vec{q},z_{\nu})\mid^{2}\right>G_{0}(\vec{k}-\vec{q},z_{\nu}-
z_{\alpha})
}=\nonumber \\
{1\over 2\pi}\sum_{\vec{q}}^{}{\int_{-\infty}^{+\infty}{
Im\left[\Xi(\vec{q},\zeta)\right]G_{0}(\vec{k}-\vec{q},\zeta-
z_{\nu})n_{B}(\zeta)d\zeta
}}-\nonumber \\
{1\over 2\pi}\sum_{\vec{q}}^{}{\int_{-\infty}^{+\infty}{\!
\Xi(\vec{q},z_{\nu}+\eta)Im\!\left[G_{0}(\vec{k}-\vec{q},\eta)
\right]\! n_{F}(\eta)d\eta}},\;\;\;
\end{eqnarray}
where $n_{B}(\omega)$ and $n_{F}(\omega)$, respectively, are the usual
Bose-Einstein and Fermi distribution
functions.\\
Contrary to the self-consistent T-matrix approach \cite{Micnas},
the self-energy given by
Eqn. (\ref{e2}) involves the non-interacting
one-particle Green's function $G_{0}(\vec{k},z)$ \cite{Ped} : on one hand,
this choice provides us with analytically
tractable expressions while, on the other hand, it should give physically
reasonable results, because approximations which include self-consistency
without the necessary vertex corrections are unable to reproduce
the precursor effects \cite{New1}, which we aim at putting in evidence.
We use the isotropic spectrum
$E(\vec{k})={k^{2}\over 2m}-\mu$.
For the values of the Ginzburg-Landau parameters $a$ and $c$,
we use the strong-coupling limit \cite{Ped,Randeria},
whereas we determine $d_{1}$ and $d_{2}$ from
the best fit to the T-matrix resulting from Monte-Carlo calculations
\cite{Ped,Micnas}.
The imaginary part of the self-energy
is shown on Fig. \ref{fig1}, for
two temperatures above the transition temperature,
which is defined by $a(T_{C})=0$. We have chosen
$T_{C}=0.1t$ ($t$ is the hopping parameter of the
original Hubbard Hamiltonian), in agreement with
T-matrix calculations \cite{Micnas}
and numerical simulations \cite{New3}.
A transition at a non-zero temperature (which is also
observed in Quantum Monte Carlo calculations, due to
finite size effects \cite{New3}) would either have to be identified
with a Berezinskii-Kosterlitz-Thouless transition or as being
due to a small coupling in the third dimension.
The most important features are the following : {\it (i)} the
imaginary part of the self-energy shows a strong dependence upon
wavenumber; upon increasing $k$, 
the rather pronounced peak moves down to the chemical potential,
crosses it at $k=k_{F}$ and then continues down to lower energies; {\it (ii)}
as the temperature increases, the imaginary part of the self-energy
presents essentially the same structure, but the peaks get
substantially broadened; {\it (iii)} for large wavenumbers, 
$-Im\left[\sigma(\vec{k},\tilde{\Omega})\right]$
becomes almost featureless; this result probably arises from
the fact that we have worked within the continuum limit while, on
the lattice, the appearance of the so-called $\eta$-resonance
would significantly modify
the spectral properties at the edge of the Brillouin zone
\cite{Kagan};
{\it (iv)} interestingly enough, we note that, except near the Fermi
wavenumber, the imaginary part of the self-energy shows a minimum
near the chemical potential, which, according to many authors \cite{Levin},
is a salient feature of a Landau Fermi liquid; however,
for $k\sim k_{F}$, we note that a rather pronounced peak appears
near the chemical potential.\\
The real part of the self-energy given by Eqn. (\ref{e2}) is computed
by performing a Kramers-Kronig transformation; thereafter, the
one-particle Green's function is obtained. In Fig. \ref{fig2}, we present the
one-particle spectral functions for two different temperatures and for
different wavenumbers. 
First of all, similarly to the case for
the imaginary part of the self-energy,
the spectral peaks broaden up, upon increasing
the temperature above $T_{C}$; this is essentially
the only effect and, in particular, we note that the position of the
peaks does not change.
Secondly, for the wavenumbers
around $k_{F}$,
the spectral function acquires a two-peak structure, which
is symmetric with respect to the chemical potential
for $k=k_{F}$. This feature persists to higher temperatures,
though the thermal
broadening tends to smear out the two-peaked structure.\\
On Fig. \ref{fig3}, we report the positions
of the spectral peaks, which have a weight significantly different
from zero. We note that, contrarily to the phase
fluctuation approach that we presented in \cite{Max1},
the opening of a pseudogap, upon approaching $T_{C}$
from above, is principally due to the double-peak structure around $k_{F}$.
\\
Summarizing, we have computed the excitation spectrum
emerging from the two-dimensional attractive Hubbard model,
upon treating the pair fluctuations above $T_{C}$
by means of a time-dependent Ginzburg-Landau equation.
Around the Fermi wavenumber,
the spectral functions exhibit a double-peak structure,
the separation of which is essentially temperature-independent.
Moreover, the spectral peaks sharpen when $T_{C}$ is approached
from above. The resulting one-particle density of states thus
shows a pseudogap of almost constant width,
above the superconducting transition temperature.
\\
This work was supported by the Swiss National Science
Foundation.

\includegraphics{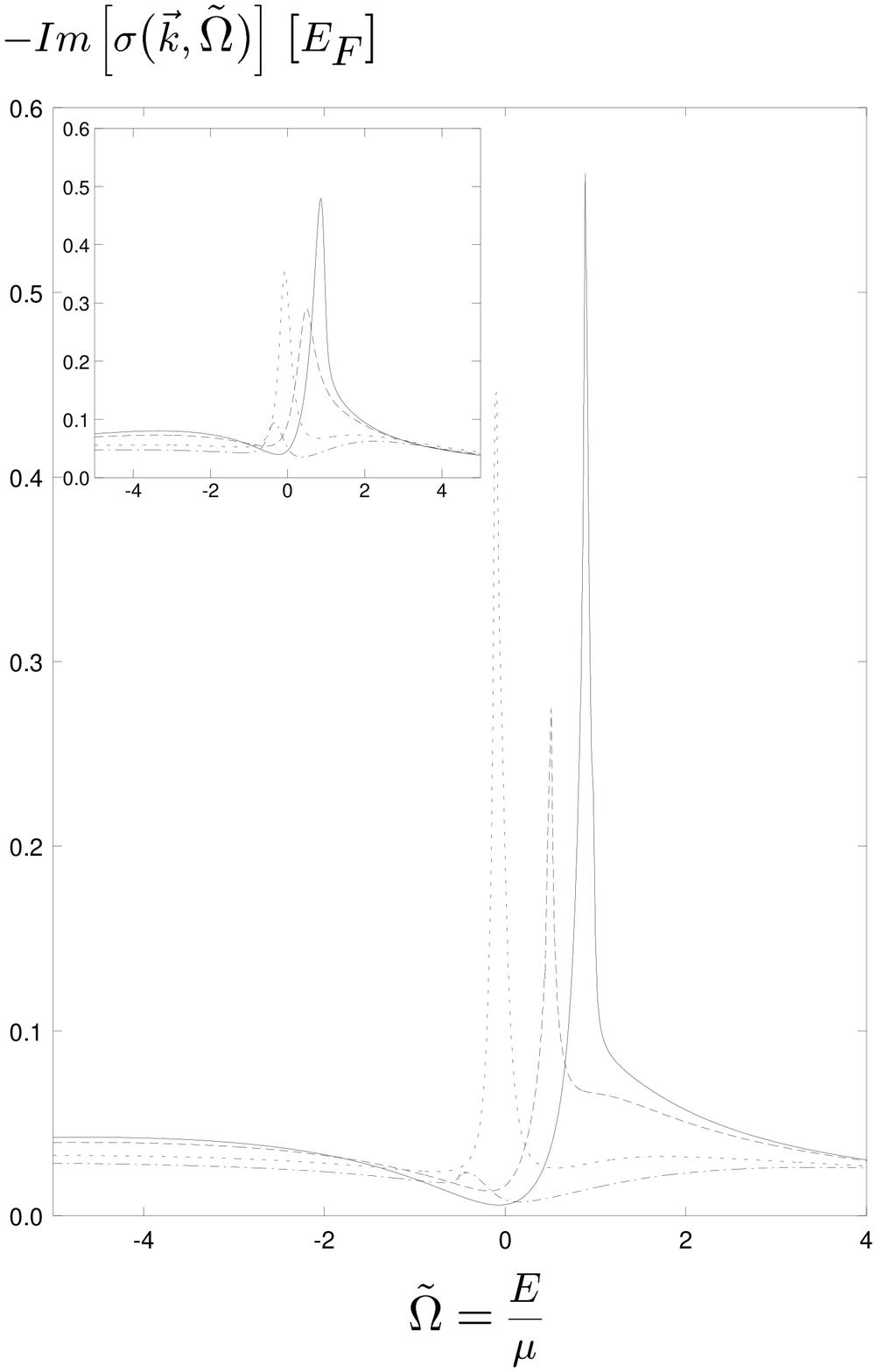}
\vglue 12.0truecm
\medskip
\begin{figure}
\caption{
Imaginary part of the self-energy, as calculated from Eqn. (\ref{e2}),
in units of the Fermi energy $E_{F}$,
for $k=0.1k_{F}$ (full line),
$k=0.5k_{F}$ (dashed line),
$k=1.1k_{F}$ (dotted line),
$k=1.5k_{F}$ (dash-dotted line), for $T=0.11t$ and, in the inset,
for $T=0.17t$. The chemical potential is the same for both temperatures
and it is taken to correspond to about 0.1 charge carriers per site.}
\label{fig1}
\end{figure}
\newpage
\includegraphics{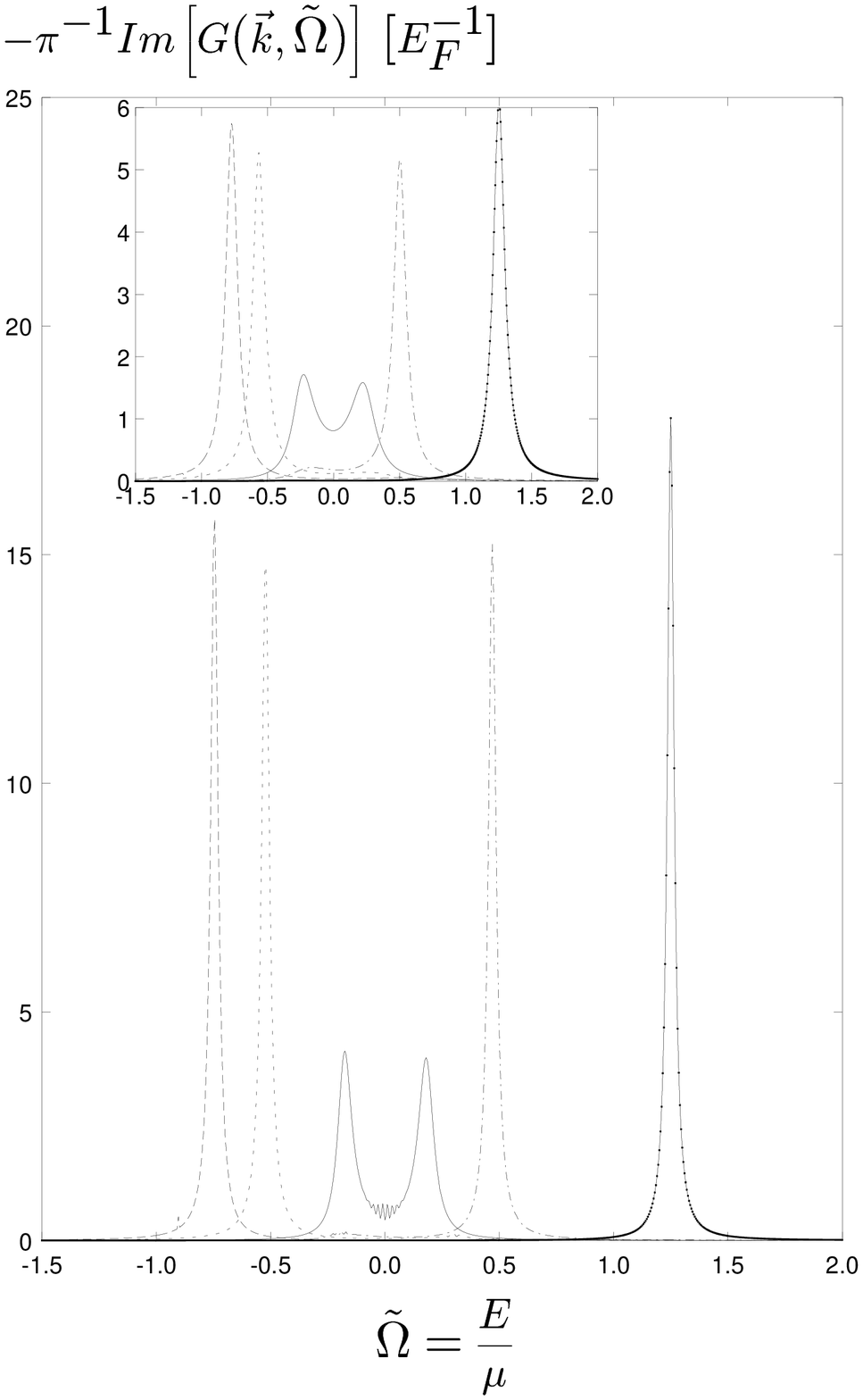}
\vglue 12.0truecm
\medskip
\begin{figure}
\caption{
One-particle spectral functions for $k=0.5k_{F}$ (dashed line),
$k=0.7k_{F}$ (dotted line),
$k=k_{F}$ (full line),
$k=1.2k_{F}$ (dash-dotted line),
$k=1.5k_{F}$ (heavy full line), for $T=0.11t$ and (inset)
$T=0.17t$.}
\label{fig2}
\end{figure}
\newpage
\newpage
\includegraphics{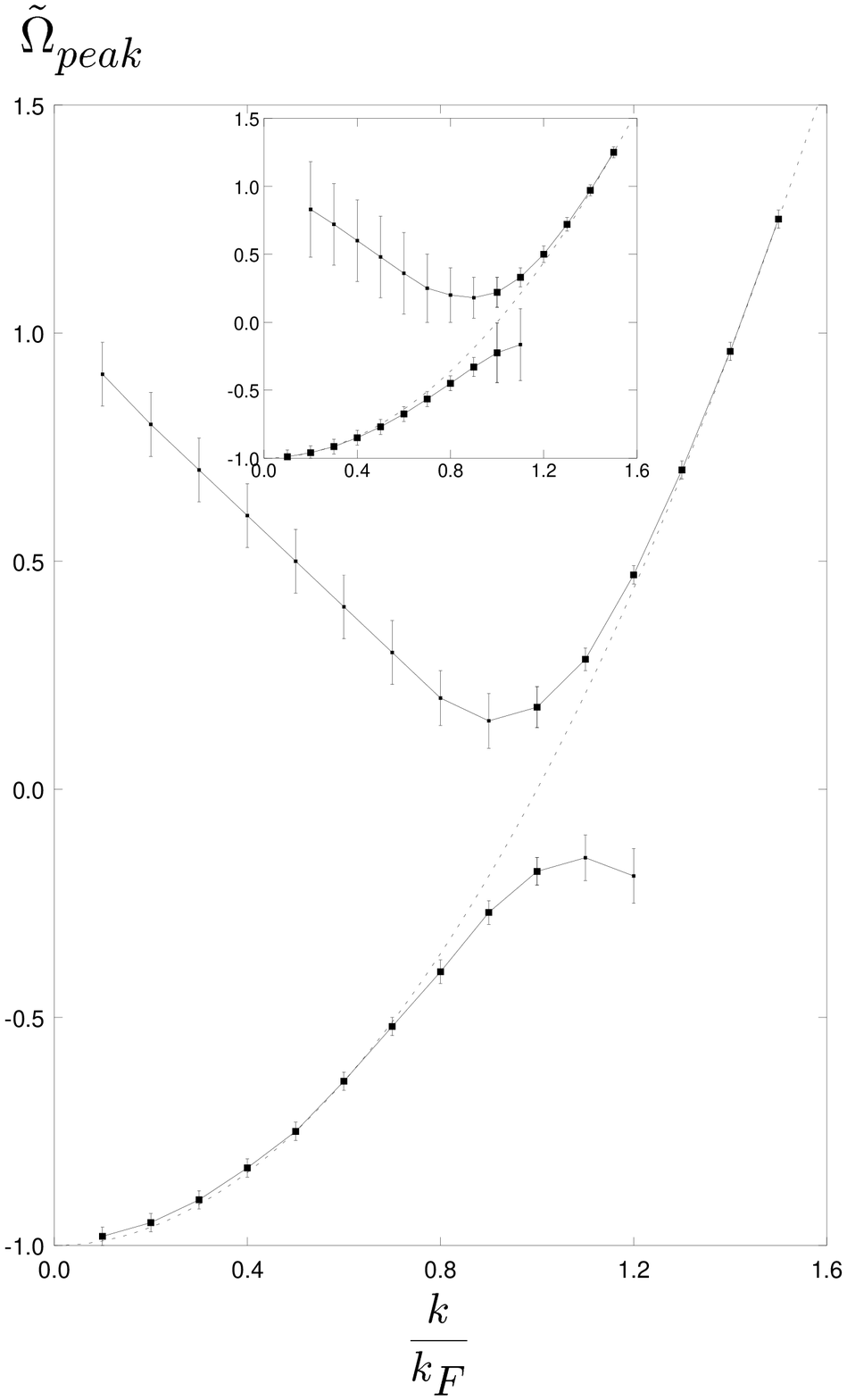}
\vglue 12.0truecm
\medskip
\begin{figure}
\caption{
Position of the spectral peaks for $T=0.11t$ and, in the inset,
for $T=0.17t$. The big squares indicate the peaks that present a
substantial spectral weight, while the small ones correspond
to peaks with very small weight.}
\label{fig3}
\end{figure}
\end{document}